\documentclass[a4paper,fleqn]{cas-sc}

\usepackage[numbers]{natbib}

\def\tsc#1{\csdef{#1}{\textsc{\lowercase{#1}}\xspace}}
\tsc{WGM}
\tsc{QE}
\tsc{EP}
\tsc{PMS}
\tsc{BEC}
\tsc{DE}

\begin{document}

\let\WriteBookmarks\relax

\title [mode = title]{A Reciprocity-Based Signal Compensation Framework for Ultrasonic Backscatter Measurements in Heterogeneous Scattering Media}                      



\author[1,2]{Wei Yi Yeoh}[type=editor]
\cormark[1]
\ead{yeohwy1@a-star.edu.sg}


\affiliation[1]{organization={Advanced Remanufacturing \& Technology Centre (ARTC), Agency for Science, Technology and Research (A*STAR)},
                addressline={3 Cleantech Loop, \#01/01 CleanTech Two}, 
                postcode={637143}, 
                country={Republic of Singapore}}

\author[2]{Bo Lan}
\author[2]{Michael J. S. Lowe}

\affiliation[2]{organization={Imperial College London},
                addressline={South Kensington}, 
                postcode={SW7 2AZ}, 
                city={London},
                country={United Kingdom}}



\cortext[cor1]{Corresponding author}


\begin{abstract}
Ultrasonic backscatter measurements are widely used for microstructural characterisation. However, in materials containing strong anisotropy and spatial heterogeneity, the interpretation of backscatter signals becomes challenging because distance-dependent propagation effects can obscure genuine microstructural variations across depth. In this paper, a cross-directional compensation method is presented for ultrasonic backscatter measurements acquired from opposing inspection surfaces. The method exploits the reciprocal constraint that the dominant through-thickness propagation bias should contain a shared component between opposing inspection directions. A shared distance-dependent baseline is estimated in the logarithmic amplitude domain using an anchor-based fitting approach and subsequently used to compensate the measured backscatter profiles with depth. The method is demonstrated on two macrozone-containing Ti--6Al--4V samples, where conventional attenuation-based compensation is shown to be insufficient to consistently reconcile opposing-face backscatter profiles. Across six opposing-face signal pairs, the proposed method reduces the mean standard deviation of the directional mismatch profile from $0.367$ to $0.120$ and the mean absolute fitted gradient from $0.171$ to $0.0067$, outperforming conventional attenuation compensation. These results demonstrate that reciprocity-based compensation can reduce propagation-related bias while preserving local direction-dependent scattering variations, providing a practical signal-normalisation framework for backscatter analysis in heterogeneous anisotropic materials.
\end{abstract}



\begin{keywords}
Wave Scattering \sep Signal Processing \sep Heterogeneous Media \sep Ultrasonic Backscatter \sep Polycrystalline Materials \sep Macrozone Characterisation \sep Volumetric Backscatter Mapping
\end{keywords}

\maketitle
\thispagestyle{empty}

\section{Introduction}

Titanium alloys such as Ti--6Al--4V (Ti--64) are widely used in critical aerospace components due to their high specific strength and excellent fatigue resistance. However, during thermomechanical processing, these alloys may develop macrozones, defined as spatially extended regions of similarly oriented crystallographic texture that are known to adversely affect fatigue behaviour under dwell loading conditions \cite{Zheng2016,Xu2020,Wu2022}. As a result, there is significant interest in non-destructive techniques capable of assessing macrozone-related heterogeneity in bulk components.

Ultrasonic inspection provides several approaches for microstructural characterisation, most commonly through measurements of attenuation, velocity, or backscatter. Attenuation measurements quantify the cumulative loss of wave energy during propagation and have been widely used to estimate grain size and scattering strength~\cite{Huang2021a,Liu2021a,ruiz2024ultrasonic}. Velocity-based approaches are correlated with the overall elastic stiffness of the material~\cite{Lan2014a} and have been used for bulk texture inversion~\cite{Bo2018}. However, both types of measurements are typically spatially averaged along the propagation path and therefore provide limited information about spatial microstructural variations.

Backscatter-based microstructure characterisation is commonly formulated through grain-noise and figure-of-merit approaches, where the measured response is related to statistical descriptions of scattering strength in polycrystalline media~\cite{Margetan1993a,Margetan1994}. This was later developed through analytical formulations, numerical methods, and array-based approaches for grain-size evaluation in metallic materials~\cite{Liu2019,Bu2021,Wang2025}. These conventional backscatter descriptors are most straightforward to interpret in statistically homogeneous materials, where propagation effects and inspection-system response can be characterised. Recent work has also extended backscatter analysis to non-uniform polycrystalline microstructures, highlighting its potential for depth-dependent microstructure assessment~\cite{Liu2025}.

However, extending depth-resolved backscatter interpretation to highly anisotropic and spatially heterogeneous materials remains challenging. A relevant example is Ti--64 containing macrozones, which are clusters of grains with preferred crystallographic orientations that may be viewed, in a simplified scattering sense, as large textured inclusions embedded within a finer polycrystalline background~\cite{Cuddihy2017,Liu2021}. As ultrasound propagates through such heterogeneous media, distance-dependent effects including beam divergence, cumulative scattering, and anisotropic wavefield distortion influence the measured backscatter amplitude~\cite{Yeoh2023}. These effects can introduce systematic propagation-related artefacts that make it challenging to distinguish genuine microstructural variations from measurement-depth bias.

A number of analytical and semi-analytical approaches have been proposed to model ultrasonic scattering from macrozones by representing the material as a homogeneous assembly of elongated domains or textured inclusions \cite{Lobkis2012,Yang2013,Pilchak2014b,Rokhlin2021a}. Such methods can provide useful bulk-averaged information through attenuation or backscatter ratio measurements, but they inherently average over spatial heterogeneity and are therefore insensitive to depth-resolved or spatially localised macrozone features. Moreover, these approaches do not readily support correction of distance-dependent bias in experimental backscatter data when the assumptions of homogeneity and weak anisotropy are violated. While numerical modelling approaches such as finite-element simulations can capture some of these effects under idealised conditions \cite{Yeoh2025}, experimental data remain essential for assessing real components, where microstructural morphology, spatial distribution, and crystal orientations are not known beforehand.

These limitations motivate the need for practical signal-processing strategies that reduce distance-related bias in ultrasonic backscatter measurements without relying on detailed microstructural models or assumptions of homogeneous material behaviour. When opposing inspection surfaces are available, the overall through-thickness transmission loss must be identical when the inspection direction is reversed. However, in heterogeneous materials the depth-resolved backscatter profiles may differ between directions because the ordering of scattering regions along the propagation path changes.

In this work, a cross-directional compensation method is proposed to reduce distance-dependent bias in ultrasonic backscatter profiles acquired from opposing inspection faces. The method exploits the reciprocal constraint on through-thickness propagation while allowing the local backscatter response to remain direction-dependent. The resulting compensation improves the directional consistency in measured backscatter amplitudes and enables clearer interpretation of spatial backscatter variations in heterogeneous anisotropic materials.

The main contributions of this work are threefold. First, experimental measurements demonstrate that reciprocal agreement in attenuation does not necessarily imply cross-directional consistency in depth-resolved backscatter profiles. Second, a log-domain reciprocity-based framework is introduced to estimate a shared depth-dependent propagation baseline directly from opposing-face backscatter measurements. Third, the improvement in bidirectional consistency is quantified using monotonic bias metrics and volumetric hotspot overlap metrics.

The remainder of this paper is organised as follows. Section~2 describes the experimental configuration and ultrasonic data used in this study. Section~3 presents the observed opposing-face backscatter behaviour and identifies the limitations of attenuation-based compensation. Section~4 introduces the reciprocity-based compensation framework and evaluates its performance and robustness. Section~5 discusses its role as a signal normalisation step prior to microstructure inference and evaluates volumetric hotspot consistency. Finally, Section~6 summarises the main findings and conclusions of the present work.

\section{Samples and Experimental Setup}

\begin{figure}
	\centering
	\includegraphics[width=1\textwidth]{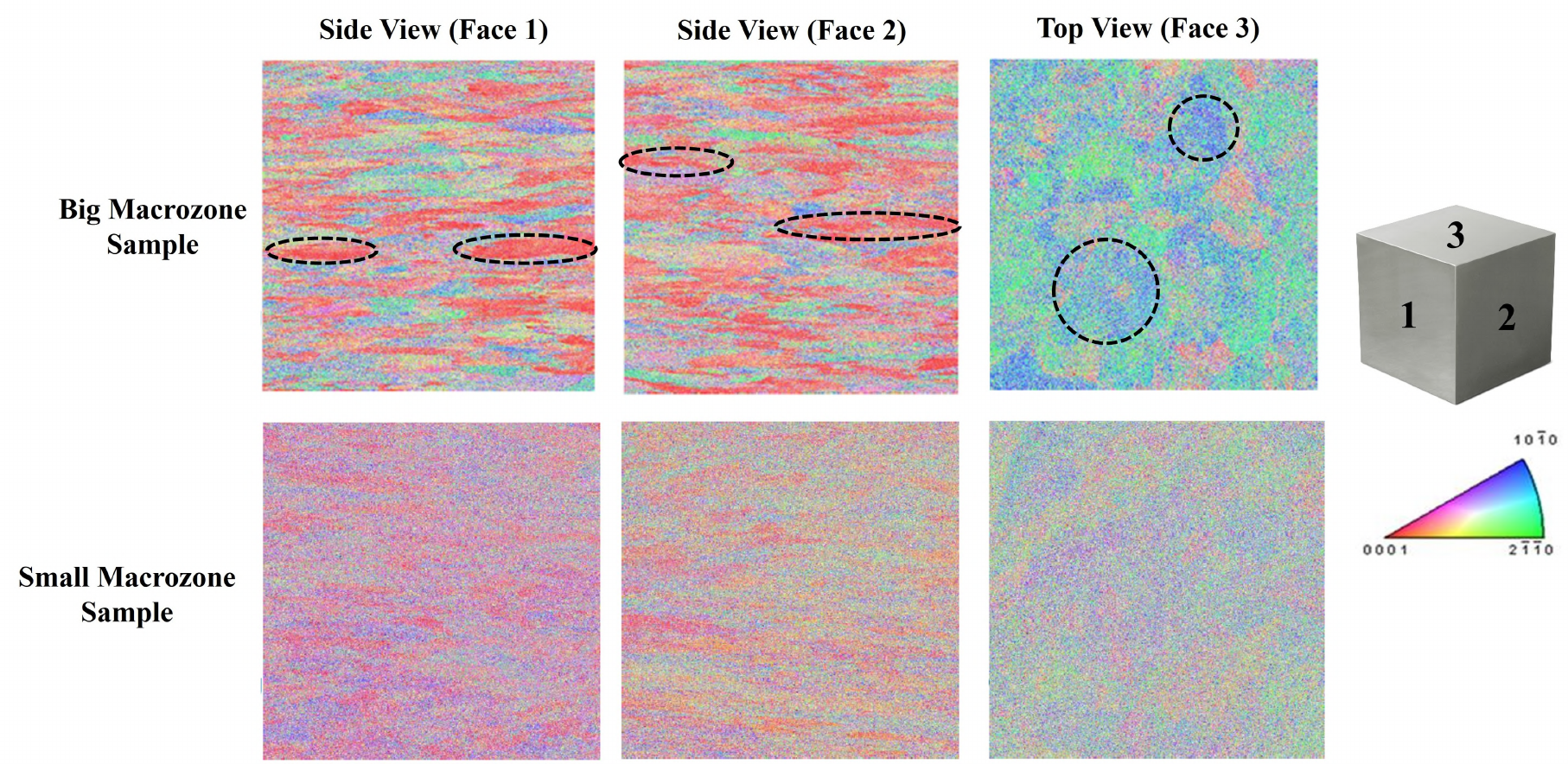}
	\caption{EBSD images of the three orthogonal surfaces of the small- and big-MTR Ti--64 cubic specimen (10 $\times$ 10 mm area). The big-MTR sample contains macrozones with pancake-like geometries that are outlined for illustration. The colour code represents the grain orientations as depicted by the Inverse Pole Figure. Adapted from \cite{Yeoh2023}.}
	\label{fig:EBSD_Ti64}
\end{figure}

In this study, two Titanium-6Al-4V (Ti--64) cubic samples with dimensions of 45 $\times$ 45 $\times$ 45 mm containing macrozones were used as representative samples for heterogeneous anisotropic scattering media. The samples were subjected to specific heat treatment and deformation processes to produce distinct macrozone distributions, as illustrated in the EBSD maps in Figure \ref{fig:EBSD_Ti64} labelled as small- and big-MTR. The EBSD maps reveal large elongated macrozones in the big-MTR sample that extend over several millimetres, characterised by clusters of similarly oriented alpha grains embedded within a finer polycrystalline background. The macrozones exhibit strong crystallographic texture with preferred orientations, resulting in spatially heterogeneous elastic anisotropy.

Ultrasonic measurements were performed using a water-immersion pulse-echo setup with a 10 MHz planar transducer (diameter: 6.35 mm, stand-off: 25 mm, step size: 0.2 mm). Preliminary measurements were also conducted at lower frequencies, but 10 MHz was selected as it provided a suitable balance between spatial resolution and penetration depth to resolve backscatter variations within the macrozone structure. A Peak NDT LT2 ultrasonic inspection system was used for signal acquisition at a sampling rate of 200 MHz, with 128 averages applied to improve the signal-to-noise ratio. Ultrasonic measurements were acquired on all 6 faces of the samples over a 120 $\times$ 120 spatial grid using the same experimental configuration described in \cite{Yeoh2023}. The dataset used in this study was collected using the same specimen and measurement setup but is analysed here specifically for the investigation of cross-directional backscatter compensation.

For general ultrasound attenuation measurements, the amplitude ratio between the front and back wall reflections is used. This ratio is typically corrected by accounting for the water-solid interface transmission and reflection coefficients, as well as a beam-diffraction compensation factor to address beam spreading effects from a finite-sized transducer \cite{Rogers1974a}. The corrected attenuation is represented by the following formula \cite{Zeng2010}
\begin{equation} \label{eq:att1}
\alpha_s = \frac{1}{z} \ln\left( \frac{FD_1 T_{01}R_{10}T_{10}}{BD_0 R_{01}} \right)
\end{equation}
where $z$ is the propagation distance, $F$ and $B$ are the front and back wall signal amplitude in the frequency domain respectively, $D$ is the beam diffraction coefficient, $T$ and $R$ are the transmission and reflection coefficients respectively, and the indices 0 and 1 refer to the water and Ti--64 media respectively.

For backscatter signals, a common approach is to construct a backscatter amplitude profile by computing the root-mean-square (RMS) amplitude across multiple spatial measurements in the time domain \cite{Margetan1993,Margetan1994}. Grain noise arises from incoherent scattering produced by random microstructural variations, resulting in strong spatial and phase variability between individual signals that are not fully representative of the underlying scattering strength. The RMS approach therefore provides a statistical measure of the average scattered energy within the sample while suppressing position-specific interference effects, yielding a more stable estimate of the scattering intensity associated with the material microstructure. In this study, prior to RMS processing, a fixed time gate was applied between the front and back wall reflections to isolate the backscatter signal while excluding potential boundary artefacts.

\section{Experimental Results}
\subsection{Observed Backscatter Behaviour}

\begin{figure}
        \includegraphics[width=.9\textwidth]{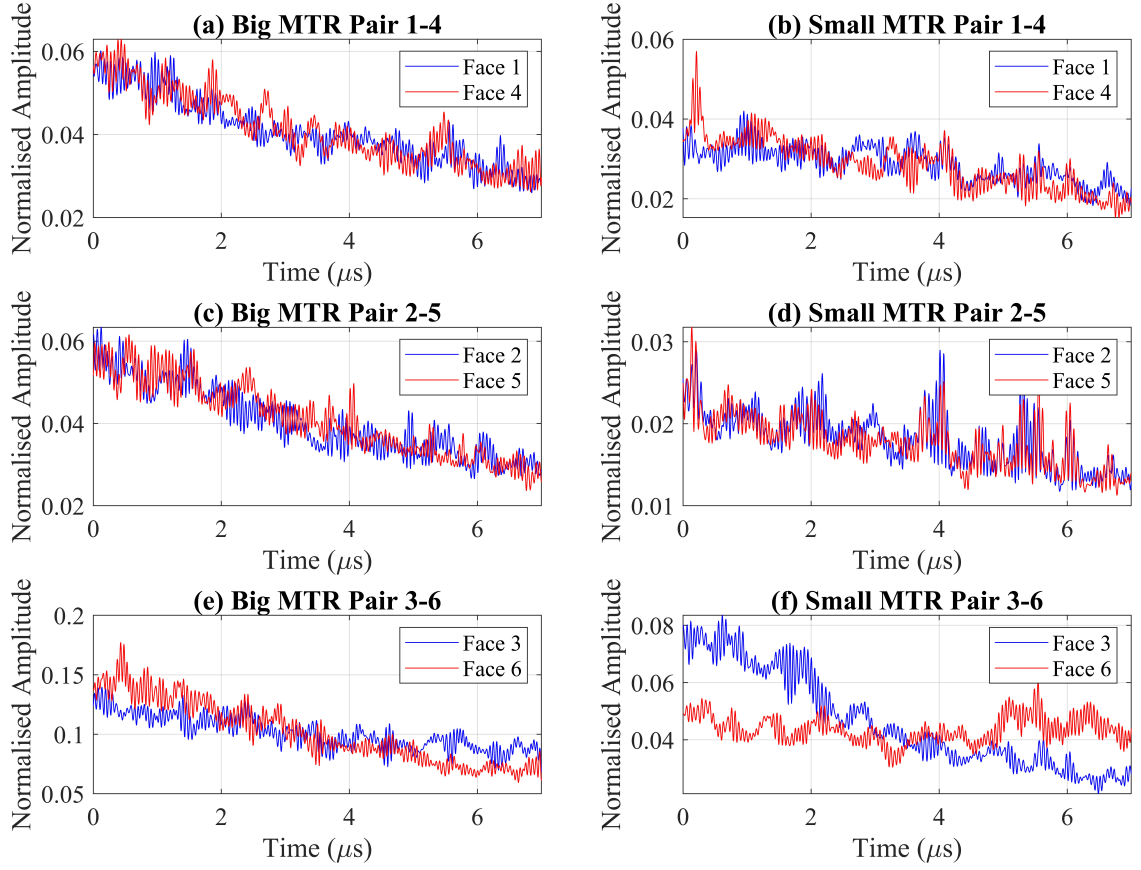}
	\centering
	\caption{Normalised backscatter amplitude profiles measured from the big- and small-MTR samples, with each subplot representing one opposing-face pair. Pairs in (a), (c), and (d) show high profile agreement, pairs in (b) and (e) show moderate agreement, and pair (f) shows low agreement based on the near-field amplitude and fitted-gradient differences.}
	\label{fig:F2}
\end{figure}

Depth-resolved ultrasonic backscatter measurements were obtained from opposing inspection faces of the two Ti--64 samples and subsequently processed to generate the RMS profiles in Figure \ref{fig:F2}. In general, regardless of sample type or inspection direction, the backscatter amplitude decreases with propagation distance due to cumulative energy losses associated with beam divergence and wave scattering. However, for cases such as (b), (e), and (f), the profiles obtained from opposing-face measurements differ significantly, while others such as (a), (c), and (d) are more comparable.

To quantify the degree of agreement between opposing-face backscatter profiles, each directional pair was characterised using two descriptors: the normalised near-field amplitude difference and the disparity between fitted log-amplitude decay gradients. For the big-MTR pair in Figure~\ref{fig:F2}(e), the profile along face 6 exhibits a higher near-field amplitude difference of approximately 1.4 $dB$ and decays at an exponential rate of $0.122\,\mu\mathrm{s}^{-1}$, which is $114\%$ faster than that of face 3 which decays at a rate of $0.057\,\mu\mathrm{s}^{-1}$. Similarly, for the small-MTR pair in Figure~\ref{fig:F2}(f), the profile of face 3 has a larger near-field amplitude difference of approximately 4 $dB$ and a fitted decay rate that is almost two orders of magnitude higher than that of face 6. These cases illustrate that directional disagreement can arise from both the initial amplitude offset and subsequent depth-dependent decay.

Even though the signal pairs are measured from the same samples, the backscatter measurements are direction-dependent. This behaviour is consistent with the macrozone structure of the Ti--64 samples, where spatial heterogeneity and elastic anisotropy influence the measured backscatter response \cite{Yeoh2023,Yang2012c}. The quantitative metrics to account for these variations are computed for all signal pairs and are listed in Table \ref{tab:bs_comparison}. Profile agreement is introduced as a qualitative descriptor to describe the similarity between opposing-face measurements based on the normalised near-field amplitude and decay-gradient differences.

\begin{table}[htbp]
\centering
\caption{Quantitative comparison of opposing-face backscatter profiles. Amplitude difference is computed as the near-field amplitude difference in dB, and decay gradients are obtained from linear fits over the selected depth window. $\Delta g$ denotes the absolute gradient difference between opposing directions.}
\label{tab:bs_comparison}
\begin{tabular}{c c c c c c c}
\toprule
Sample & Face Pair (A-B) & Amp Diff (dB) & Grad A & Grad B & $\Delta g$ & Profile Agreement \\
\midrule
Big   & 1--4 & 0.115 & 0.090 & 0.094 & 0.005 & High \\
Big   & 2--5 & 0.061 & 0.095 & 0.104 & 0.008 & High \\
Big   & 3--6 & 1.425 & 0.057 & 0.122 & 0.065 & Moderate \\

Small & 1--4 & 1.016 & 0.057 & 0.097 & 0.040 & Moderate \\
Small & 2--5 & 0.091 & 0.068 & 0.067 & 0.001 & High \\
Small & 3--6 & 4.000 & 0.171 & 0.002 & 0.169 & Low \\
\bottomrule
\end{tabular}
\end{table}

\subsection{Attenuation Measurements and Compensation}

\begin{figure}
        \includegraphics[width=.9\textwidth]{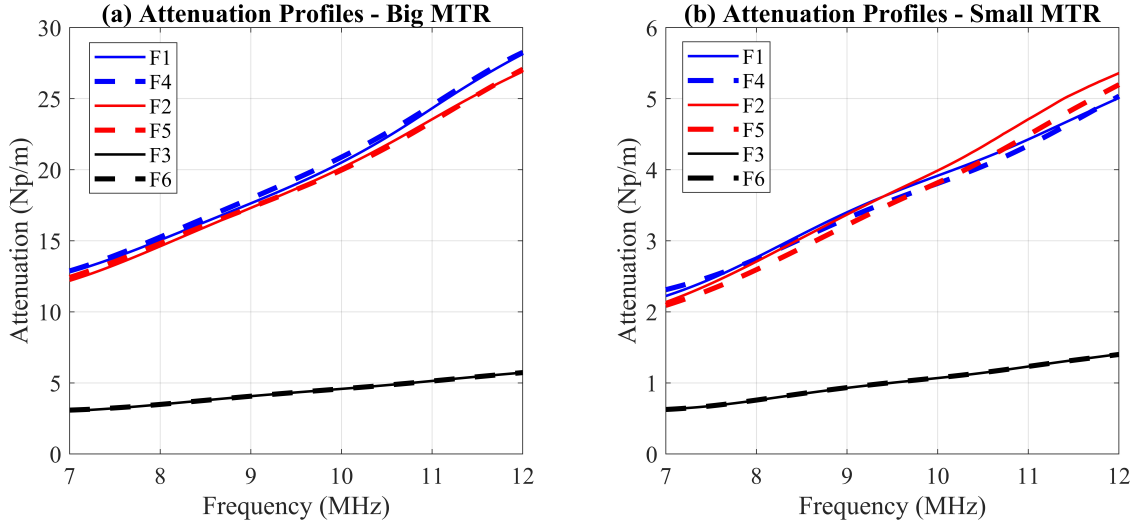}
	\centering
	\caption{Frequency-dependent attenuation profiles for the (a) big- and (b) small-MTR samples, measured from all six faces. Solid and dashed lines indicate opposing faces within each face pair.}
	\label{fig:F3}
\end{figure}

Conventionally, attenuation is often used as a compensation factor to account for propagation-related energy losses, including absorption, scattering loss from the coherent beam, beam distortion, and interface effects. In this work, attenuation profiles were estimated for all six inspection faces on the two Ti--64 samples using equation~(\ref{eq:att1}), with the resulting values averaged over the 120 $\times$ 120 spatial grid. The frequency-dependent attenuation profiles are shown in Figure~\ref{fig:F3}.

The big-MTR sample exhibits substantially higher attenuation than the small-MTR sample, particularly for face pairs 1--4 and 2--5. This is consistent with the presence of larger macrozones that introduce stronger anisotropic propagation effects and therefore higher attenuation. Directional variation is also observed between the different face pairs, reflecting the directional dependence of attenuation due to the underlying anisotropic microstructure, as evident in a prior numerical study using realistic synthetic macrozones \cite{Yeoh2025}. 

Regardless of the directional and sample variations, the attenuation profiles of opposing faces within each pair show very good agreement. This behaviour is consistent with reciprocity in linear elastic wave propagation, where for two fixed surfaces, the total transmission loss along a given path should be symmetric when the propagation direction is reversed. The agreement between opposing-face attenuation profiles therefore provides a useful baseline for conventional attenuation-based compensation.

\begin{figure}
        \includegraphics[width=.9\textwidth]{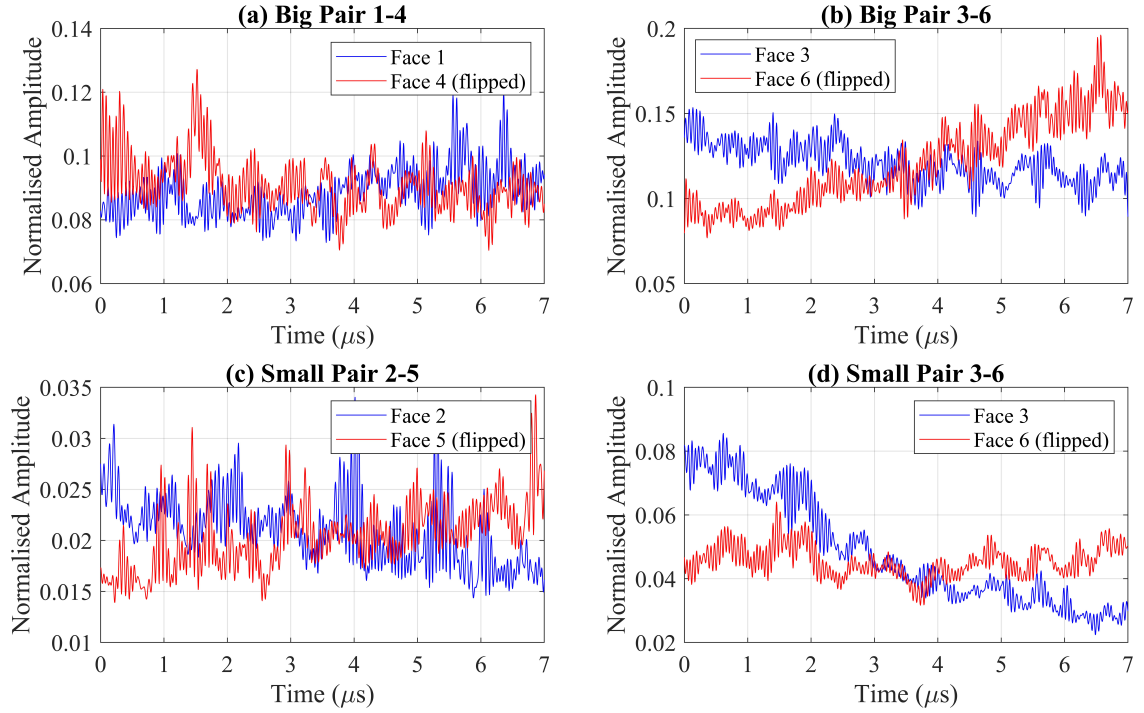}
	\centering
	\caption{Attenuation-compensated backscatter RMS profiles for selected opposing-face pairs, with the second profile flipped into a common depth coordinate. Pairs with originally high profile agreement are shown in (a) and (c), while (b) and (d) have moderate and weak agreement, respectively.}
	\label{fig:F4}
\end{figure}

The attenuation estimates in Figure~\ref{fig:F3} provide the conventional reference to compensate for depth-dependent loss in the backscatter profiles. For each inspection direction, the measured backscatter RMS profiles shown in Figure \ref{fig:F2} were corrected using the corresponding attenuation estimate at the central operating frequency of 10 MHz. The attenuation-compensated profiles for selected cases are illustrated in Figure \ref{fig:F4}, with the second profile in each pair reversed into a common depth coordinate for direct comparison.

Signal pairs with high original profile agreement, such as those in
Figure~\ref{fig:F4}(a) and (c), remain reasonably consistent after
attenuation compensation. This indicates that the method can be adequate
when opposing-face responses already have similar amplitude levels and
decay behaviour. Nevertheless, the compensation performance is not uniform, with the pair in Figure~\ref{fig:F4}(a) being slightly over-compensated while the pair in Figure~\ref{fig:F4}(c) remains under-compensated. Moreover, for pairs with stronger directional mismatch, such as those in Figure~\ref{fig:F4}(b) and (d), attenuation compensation is less effective, with significant opposing depth trends and amplitude differences still present after the correction.

These results show that reciprocal agreement in attenuation does not necessarily translate into reciprocal agreement in backscatter, and that attenuation-based compensation does not consistently reconcile opposing-face backscatter profiles. Although attenuation provides a physically meaningful measure of coherent-wave energy loss, it is not a complete descriptor of the depth-dependent variation in backscatter.

\subsection{Identified Limitations of Attenuation-Based Compensation}

In a statistically homogeneous scattering medium, a suitable propagation compensation would be expected to reduce the dominant depth dependence of the spatially averaged backscatter response, thereby improving consistency between opposing-face profiles. The persistence of residual depth-dependent trends after attenuation compensation therefore suggests that coherent-wall-echo attenuation does not fully represent the effective loss governing distributed backscatter. This limitation is particularly relevant in highly anisotropic and heterogeneous materials, where the spatial ordering, orientation, and scattering strength of microstructural features influence the generation and redistribution of backscattered energy, as discussed in greater detail below.

\subsubsection{Case 1: Distance-Dependent Directional Mismatch}

\begin{figure}
        \includegraphics[width=.9\textwidth]{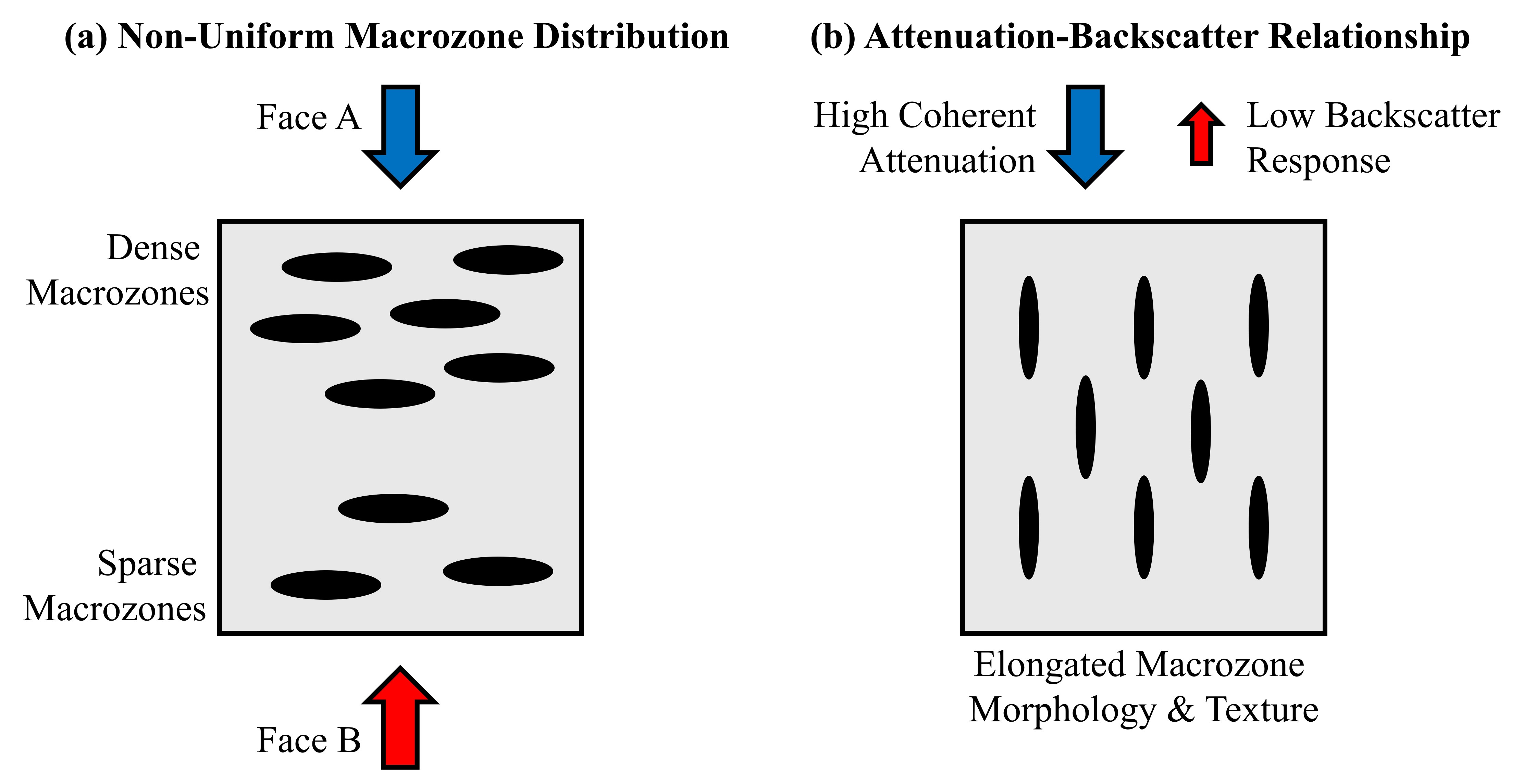}
	\centering
    \caption{Schematic illustration of two limitations of attenuation-based backscatter compensation in a material containing macrozones. (a) A non-uniform macrozone distribution causes opposing interrogation directions to encounter scattering regions in different orders, leading to direction-dependent backscatter profiles. (b) Elongated textured regions aligned with the propagation direction may produce coherent-wave distortion or attenuation without generating a proportional pulse-echo backscatter response.}
	\label{fig:F5}
\end{figure}

The first limitation occurs when opposing-face backscatter profiles exhibit clear directional mismatch. Examples are shown in Figure~\ref{fig:F2}(e) and (f) for the face 3--6 pairs in both samples. Although the profiles are obtained from opposite sides of the same sample volume, they exhibit different near-surface amplitude levels and decay trends. One plausible mechanism is the reversed ordering of heterogeneous macrozone regions along the propagation path, as illustrated schematically in Figure~\ref{fig:F5}(a). Recent work on non-uniform polycrystalline microstructures has shown that depth-varying microstructural statistics can produce depth-dependent backscatter responses~\cite{Liu2025}. In the present macrozone samples, this effect is further complicated by opposing-face inspection, where the same heterogeneous scattering regions are encountered in reverse order. As a result, propagation from one face may encounter stronger scattering regions near the entry surface, producing a larger initial backscatter response and a stronger apparent decay, whereas the reversed direction may exhibit a different depth-dependent trend.

Generally, when the interrogation direction is reversed, the wave encounters the same heterogeneous regions in the opposite order. In the absence of propagation loss and direction-dependent scattering redistribution, one might expect opposing inspections of the same volume to provide comparable depth-resolved backscatter profiles. However, in reality, the resulting backscatter profile is influenced by both the cumulative propagation loss and the depth-dependent ordering of local scattering regions along the propagation path. Hence, a scalar attenuation correction derived from wall reflections cannot fully represent this direction-dependent redistribution of backscattered energy. Consequently, attenuation compensation may reduce part of the monotonic decay, but a residual mismatch can remain between opposing depth-resolved backscatter profiles.

\subsubsection{Case 2: Inadequacy of Attenuation as a Proxy for Backscatter Propagation Losses}

The second limitation is observed in cases where opposing-face backscatter profiles already show relatively high profile agreement, such as those in Figure~\ref{fig:F2}(a), (c), and (d). In such cases, attenuation-based compensation may appear adequate because the two profiles are similar in amplitude level and decay trend and will be compensated with a similar attenuation value. However, this apparent profile agreement does not necessarily imply that the attenuation estimate provides a physically appropriate normalisation for the depth-dependent behaviour of backscatter.

This non-equivalence arises because attenuation and backscatter describe related but distinct aspects of wave interaction with the microstructure. Wall-echo attenuation primarily characterises the cumulative loss of coherent wave energy along the propagation path, including absorption, scattering out of the coherent beam, diffraction effects, beam distortion, and phase aberration associated with anisotropic texture. In contrast, pulse-echo backscatter depends on the portion of locally scattered energy redirected back toward the transducer, and is therefore influenced by the orientation, morphology, scattering strength, and spatial distribution of microstructural regions. This distinction is consistent with previous experimental and numerical work on macrozoned Ti--64, where attenuation and backscatter were shown to respond differently to macrozone morphology, texture, and scattering geometry~\cite{Yeoh2023}.

This is conceptually illustrated in Figure~\ref{fig:F5}(b). Propagation through elongated textured macrozones may produce substantial coherent-beam distortion or phase variation at the receiver, leading to an elevated attenuation estimate. However, this does not necessarily correspond to a proportional increase in backscattered energy. If the microstructural arrangement redirects only a limited fraction of energy back toward the receiver, the measured backscatter response may remain comparatively weak even when the inferred attenuation is high. Hence, direct attenuation-based compensation can over-correct the profile, as illustrated in Figure \ref{fig:F4}(a) with the compensated backscatter amplitude increasing with depth. Conversely, a low attenuation estimate does not imply that the distributed backscatter response is free from depth-dependent loss, and may instead under-compensate the measured RMS profile. Therefore, attenuation-based compensation can reduce propagation-related coherent-wave decay, but it cannot fully account for the local and directional redistribution processes that govern backscatter in anisotropic heterogeneous materials.

These cases show that attenuation-based compensation can be limited both when opposing-face profiles exhibit strong distance-dependent mismatch and when they appear superficially consistent. The limitation is that attenuation measures coherent-wave loss rather than the effective propagation trend governing distributed backscatter. This motivates a compensation strategy that uses the paired opposing-face measurements directly, leading to the reciprocity-based framework introduced in the following section.

\section{Reciprocity-Based Compensation Framework}

The compensation framework was motivated by the systematic observation that attenuation-based compensation does not consistently reconcile opposing-face backscatter profiles across the measured face pairs. The initial focus of the study was the big-MTR face 3--6 pair, which exhibits the strongest backscatter response and therefore provides the clearest high-scattering regions for the subsequent volumetric hotspot analysis. Nevertheless, the proposed compensation is formulated as a signal-level normalisation procedure and is not specific to this pair.

\begin{figure}
        \includegraphics[width=50mm]{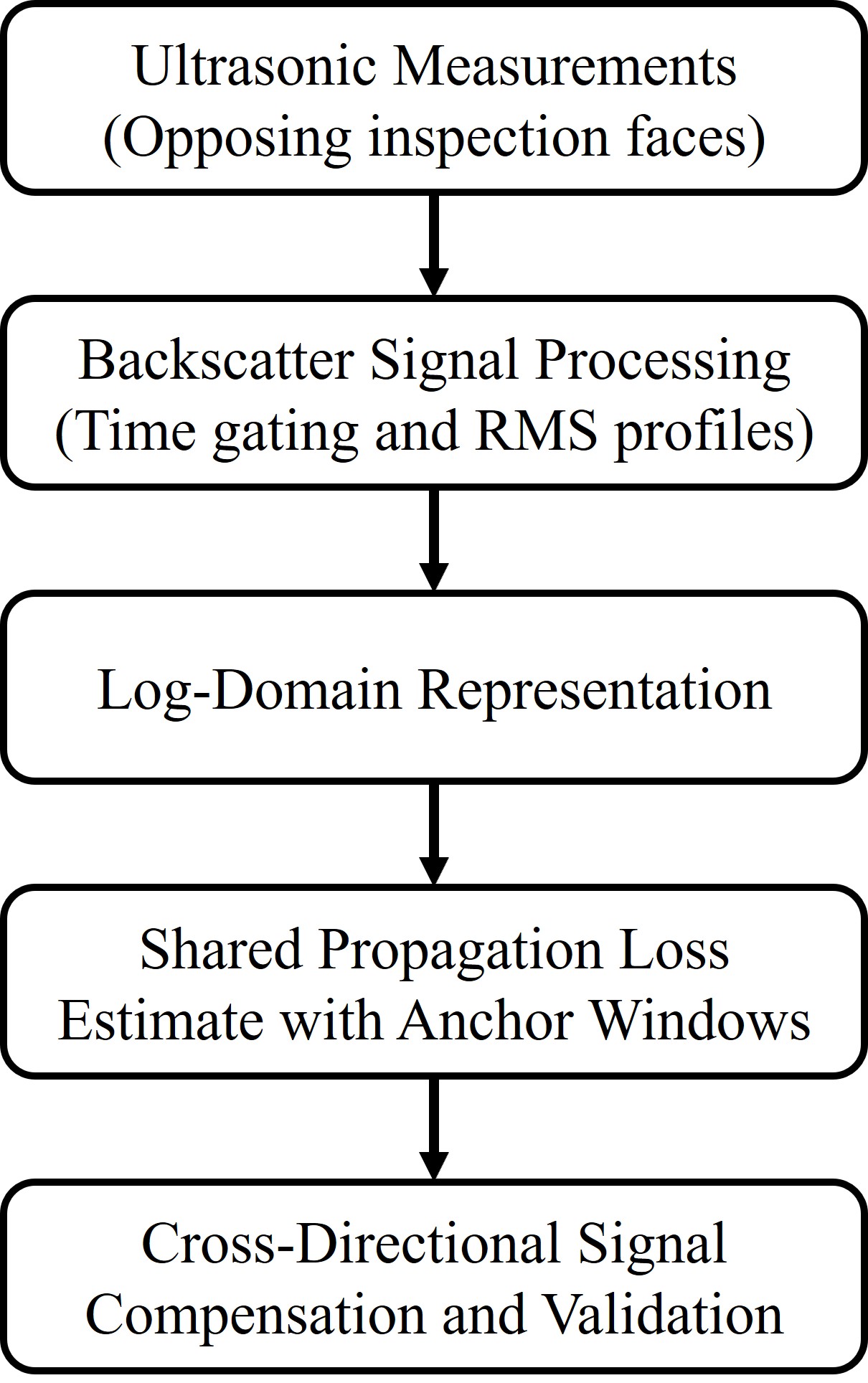}
	\centering
	\caption{A workflow that outlines the methodology for the proposed cross-compensation strategy to reduce propagation-related bias in backscatter measurements.}
	\label{fig:F6}
\end{figure}

The proposed reciprocity-based cross-directional compensation framework is outlined in Figure~\ref{fig:F6}. The method estimates a smooth depth-dependent propagation baseline directly from paired RMS backscatter profiles acquired from opposite faces of the same volume. It does not assume that the local backscatter response is identical in the two directions or that the estimated baseline represents a unique physical energy-loss profile. Instead, it assumes that the dominant through-thickness propagation bias can be approximated by a shared smooth trend when the same material volume is inspected in reverse directions. This trend is estimated in the log domain and is used to reduce depth-related amplitude bias while preserving residual local and direction-dependent scattering variations associated with microstructure heterogeneity.

\subsection{Signal Decomposition and Compensation Strategy}

The method exploits backscatter measurements from opposing inspection faces of the same component. Although absolute amplitudes may differ due to non-uniform macrozone distributions and directional scattering, propagation-related losses along the common path are expected to produce a shared depth-dependent trend. The compensation therefore targets this shared component while preserving direction-specific variations. Let $R_3(z)$ and $R_6(z)$ denote the depth-resolved RMS backscatter amplitudes from faces 3 and 6, respectively, with $z$ measured from the face 3 surface. The face 6 profile is mapped into the same coordinate system, and the analysis is performed in the logarithmic domain,
\begin{equation}
f_3(z) = \log R_3(z), \qquad
f_6(z) = \log R_6(z),
\end{equation}
where multiplicative propagation effects become additive. The profiles are then modelled as a shared depth-dependent loss term plus a direction-specific scattering residual term. In this representation, the propagation-related contribution typically appears as a slowly varying trend with depth, while directional scattering variations manifest as residual fluctuations. The shared baseline loss function $\ell(z)$ is therefore represented by
\begin{equation}
f_i(z) = s_i(z) + \ell(z), \qquad i \in \{3,6\},
\end{equation}
where $s_i(z)$ represents direction-dependent scattering contributions that are not explicitly modelled. 

The logarithmic form of the RMS amplitude profiles typically exhibits an approximately linear decrease with propagation distance in highly scattering materials. Hence, a minimal log-linear form is adopted for the shared loss function, 
\begin{equation}
\ell(z) = \alpha z + \beta,
\end{equation}
where $\alpha$ is an effective depth-loss rate that captures the combined influence of attenuation, beam spreading and cumulative scattering, and $\beta$ represents a constant offset associated with the reference amplitude level near the inspection surface.

The parameters of the baseline loss model are estimated using two reference windows located within the interior scattering region of the signal: a near-field window $W_n$ and a far-field window $W_f$. These windows are positioned to avoid interference from the front-wall reflection and back-wall echo while sampling regions where the RMS backscatter amplitude exhibits relatively stable behaviour. The reference levels are defined as the average logarithmic RMS amplitudes across both inspection directions,
\begin{equation}
\bar f_n = \frac{1}{2}\left( \langle f_3 \rangle_{W_n} + \langle f_6 \rangle_{W_n} \right), \qquad
\bar f_f = \frac{1}{2}\left( \langle f_3 \rangle_{W_f} + \langle f_6 \rangle_{W_f} \right),
\end{equation}
where $\langle \cdot \rangle_W$ denotes averaging over the specified window. Using the average of both inspection directions ensures that the estimated loss trend is consistent with reciprocal through-thickness transmission and reduces sensitivity to direction-specific fluctuations.

The effective depth-loss rate is then obtained from the difference between the two reference levels,
\begin{equation}
\alpha = \frac{\bar f_f - \bar f_n}{z_f - z_n},
\end{equation}
where $z_n$ and $z_f$ denote the central depths of the near-field and far-field windows. This slope represents the effective propagation loss associated with cumulative effects such as beam spreading, scattering attenuation and wavefield distortion.

A depth-dependent compensation factor is subsequently defined as
\begin{equation}
C(z) = \exp\!\left[-\bigl(\ell(z) - \ell(z_n)\bigr)\right],
\end{equation}
such that $C(z_n)=1$ at the near-field reference depth. The compensation factor is applied uniformly to both backscatter profiles,
\begin{equation}
R_i^{\mathrm{comp}}(z) = R_i(z)\,C(z), \qquad i \in \{3,6\}.
\end{equation}

\begin{figure}
        \includegraphics[width=.9\textwidth]{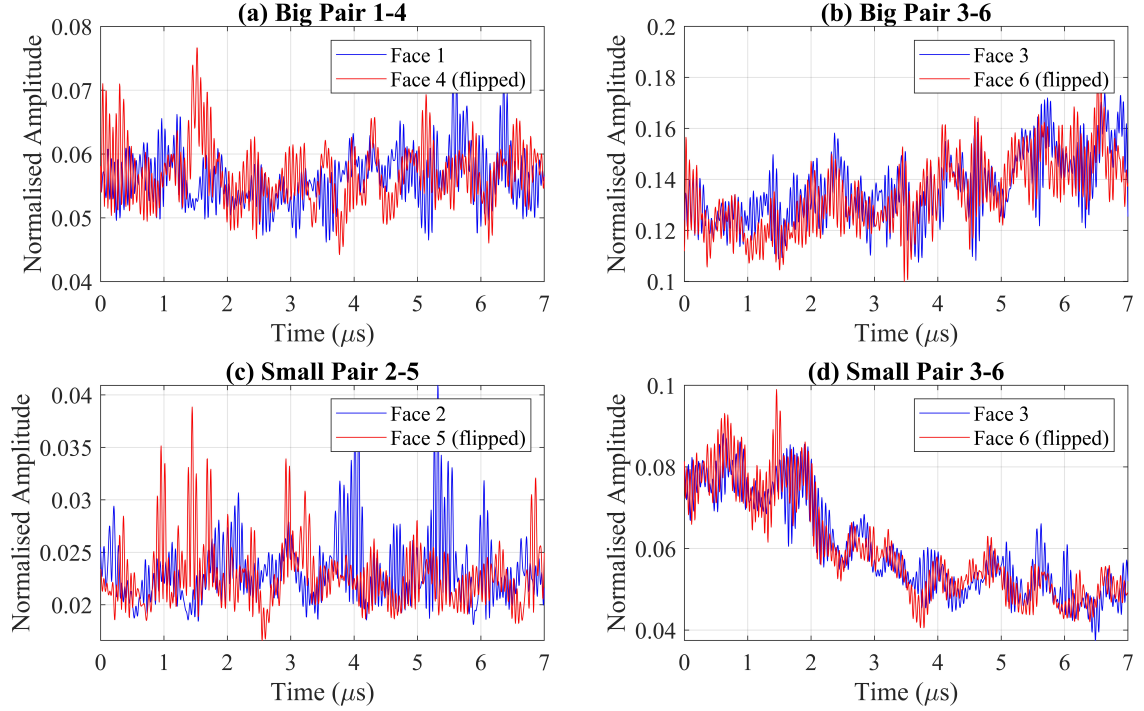}
	\centering
	\caption{Cross-compensated backscatter RMS profiles for selected opposing-face pairs, with the second profile flipped into a common depth coordinate. Pairs with originally high profile agreement are shown in (a) and (c), while (b) and (d) have moderate and weak agreement, respectively.}
	\label{fig:F7}
\end{figure}

To assess whether the method reduces the targeted monotonic bias beyond the initial big-MTR face 3--6 case, the same compensation procedure and windowing choices were applied to all six opposing-face pairs without pair-specific tuning, as shown in Figure~\ref{fig:F7}. Monotonic depth-dependent bias is substantially reduced while residual fluctuations associated with directional scattering variations remain visible. Compared to the original (Figure~\ref{fig:F2}) and attenuation-compensated (Figure~\ref{fig:F4}) backscatter signals, the signal pairs after cross-compensation exhibit stronger consistency in terms of near-surface amplitudes and profile gradients, regardless of their original degree of profile alignment.

\subsection{Reduction of Monotonic Directional Bias}

\begin{figure}
        \includegraphics[width=.9\textwidth]{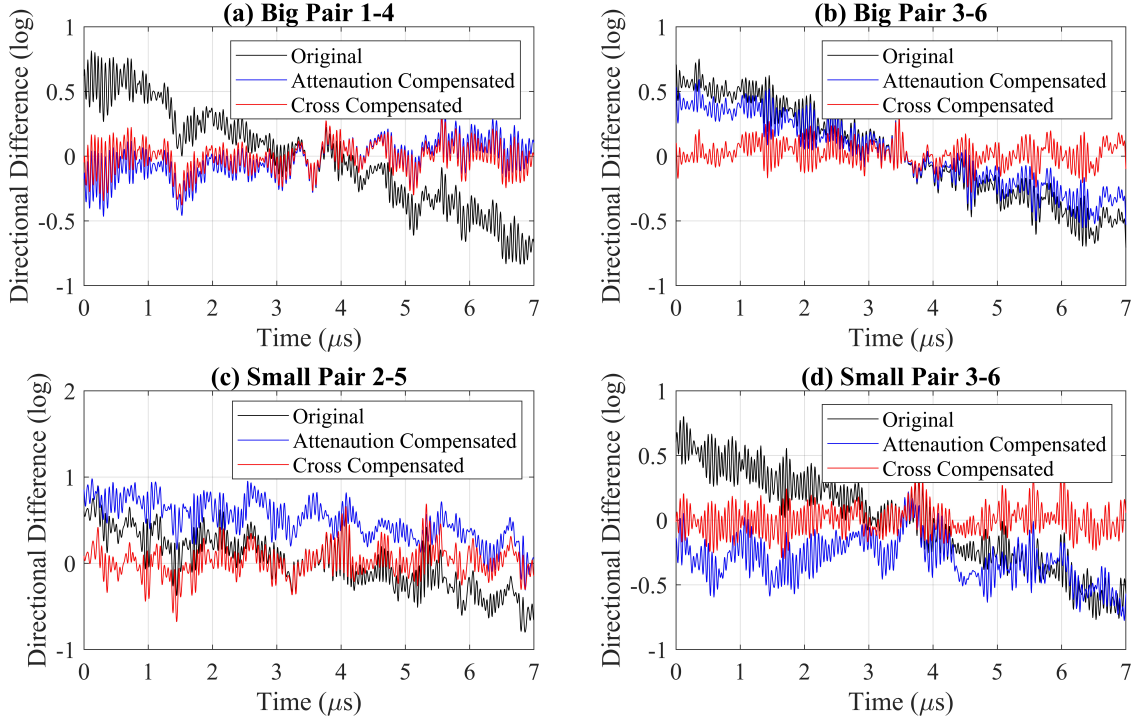}
	\centering
	\caption{Directional difference between logarithmic RMS backscatter profiles for selected opposing-face pairs, comparing the original, attenuation-compensated, and cross-compensated cases. Pairs with originally high profile agreement are shown in (a) and (c), while (b) and (d) have moderate and weak agreement, respectively.}
	\label{fig:F8}
\end{figure}

To assess the effectiveness of the proposed compensation in reducing depth-related bias, the difference between logarithmic backscatter profiles obtained from opposing inspection faces was examined as a function of depth. Specifically, with $D$ denoting the total thickness, the difference
\[
\Delta(z) = \log R_3(z) - \log R_6(D-z)
\]
is computed to quantify the residuals between opposing measurements before and after applying the compensation methods. The results are illustrated in Figure \ref{fig:F8}, with the original uncompensated cases showing monotonic trends across depth as energy losses due to beam spread and material scattering are unaccounted for. Conventional attenuation compensation partially reduces this depth-dependent trend, but a residual slope remains. In contrast, the cross-directional compensation substantially suppresses this monotonic bias, with the residual difference fluctuating about an approximately zero mean across the measured depths.

To compare the three normalisation strategies quantitatively, two metrics are extracted from the mismatch profile $\Delta(z)$: the standard deviation, which measures the overall residual variability, and the absolute fitted gradient, which measures the remaining monotonic depth-dependent bias, as listed in Table~\ref{tab:delZ}. The gradient is reported in absolute value for the comparison because positive and negative residual slopes both represent incomplete removal of depth-dependent bias.

Across the six opposing-face signal pairs, the original profiles exhibit a mean standard deviation of $0.367$ and a mean absolute fitted gradient of $0.171$, indicating a strong and systematic depth-dependent directional mismatch. Conventional attenuation compensation reduces these values to $0.199$ and $0.069$, corresponding to reductions of approximately $46\%$ and $59\%$, respectively. This confirms that attenuation compensation removes part of the propagation-related depth trend. However, noticeable residual gradients remain for several pairs, particularly Big 3--6 and Small 1--4, indicating that the coherent-wall-echo attenuation estimate does not fully describe the effective loss governing the distributed backscatter response.

In comparison, the proposed cross-directional compensation further reduces the mean standard deviation to $0.120$ and the mean absolute gradient to $0.0067$. Relative to the original profiles, this corresponds to reductions of approximately $67\%$ in variability and $96\%$ in monotonic gradient. Relative to attenuation compensation, the cross-directional method provides an additional reduction of approximately $39\%$ in variability and $90\%$ in mean absolute gradient. The much larger reduction in gradient than in standard deviation indicates that the cross-directional compensation primarily suppresses the global monotonic propagation bias, while preserving residual local and direction-dependent scattering variations associated with the macrozones in the microstructure.

\begin{table}[htbp]
\centering
\caption{Summary of monotonic depth-trend metrics for the original, attenuation-compensated, and cross-directional compensated backscatter profiles. The final row reports the mean values across all six opposing-face signal pairs.}
\label{tab:delZ}
\begin{tabular}{c c c c c c c c}
\toprule
Sample & Pair & 
\multicolumn{3}{c}{Standard Deviation} & 
\multicolumn{3}{c}{Absolute Gradient} \\
\cmidrule(lr){3-5} \cmidrule(lr){6-8}
 &  & Orig. & Att. & Cross & Orig. & Att. & Cross \\
\midrule
Big   & 1--4 & 0.388 & 0.137 & 0.111 & 0.1838 & 0.0405 & 0.0077 \\
Big   & 2--5 & 0.418 & 0.122 & 0.117 & 0.1987 & 0.0195 & 0.0091 \\
Big   & 3--6 & 0.372 & 0.274 & 0.087 & 0.1786 & 0.1283 & 0.0053 \\
Small & 1--4 & 0.336 & 0.259 & 0.126 & 0.1543 & 0.1122 & 0.0032 \\
Small & 2--5 & 0.325 & 0.228 & 0.175 & 0.1359 & 0.0837 & 0.0111 \\
Small & 3--6 & 0.365 & 0.172 & 0.106 & 0.1726 & 0.0322 & 0.0036 \\
\midrule
\multicolumn{2}{c}{Mean}
& 0.367 & 0.199 & 0.120
& 0.1707 & 0.0694 & 0.0067 \\
\bottomrule
\end{tabular}
\end{table}

The metrics show that the proposed normalisation consistently reduces the targeted monotonic bias across all six opposing-face pairs, including both big- and small-MTR samples, without pair-specific tuning. However, because all measurements come from the same experimental sample set and inspection configuration, the results should be interpreted as a specimen-level transferability assessment rather than exhaustive validation across broader heterogeneous materials or measurement geometries.

\subsection{Robustness to Window Selection}

Two robustness checks were performed to assess whether the proposed compensation framework is sensitive to windowing choices. The first evaluates the stability of the profile-agreement descriptors with respect to the isolated backscatter window length, while the second evaluates the stability of the estimated cross-compensation slope with respect to the length of the anchor-window placement.

\begin{figure}
    \includegraphics[width=.9\textwidth]{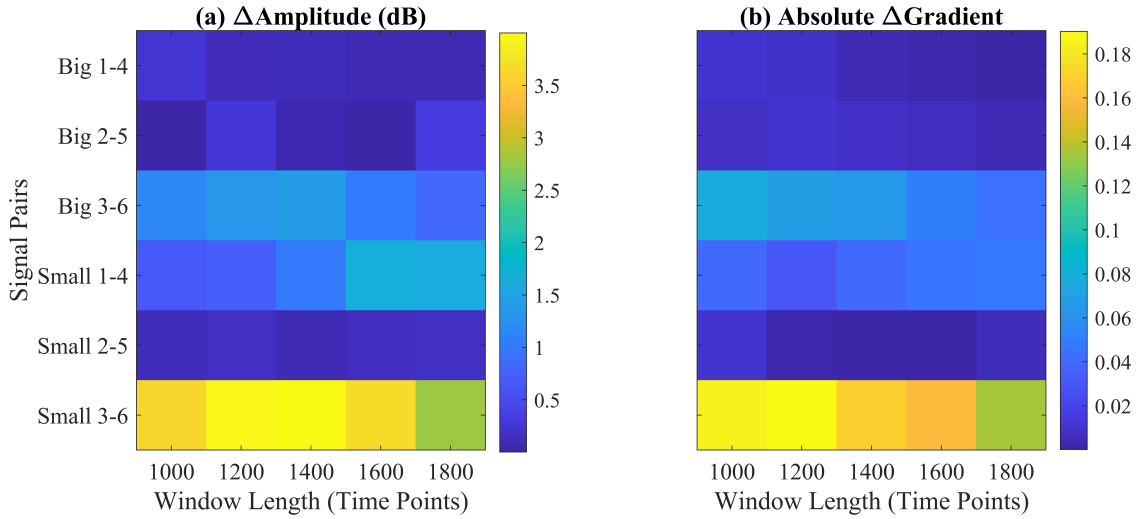}
	\centering
	\caption{Heatmaps showing the variation of (a) amplitude difference and (b) absolute fitted-gradient difference for the six opposing-face signal pairs when the backscatter window length is varied from 1000 to 1800 time points.}
	\label{fig:F9}
\end{figure}

The effect of backscatter window length is shown in Figure~\ref{fig:F9}. The RMS window length was varied from 1000 to 1800 time points, with previous results reported above computed using a window length of 1400 time points. The normalised amplitude difference and absolute fitted-gradient difference were recomputed for all six opposing-face pairs. Across the tested range, the maximum absolute changes are small, approximately $0.13 dB$ for the amplitude difference and $0.07$ for the fitted-gradient difference. Although the relative percentage variation can appear larger for cases with near-zero descriptor values, the ordering of the six
signal pairs remains unchanged for all tested window lengths. The high-agreement, moderate-agreement, and low-agreement cases therefore remain consistently classified, indicating that the profile-agreement assessment is not dominated by the backscatter window-length choice.

A separate robustness analysis was performed for the anchor-window selection used in the cross-directional compensation to evaluate the stability of the estimated compensation slope. The sensitivity of the cross-compensated monotonic bias metric to anchor-window length is summarised in Table~\ref{tab:anchor_window_length}. The near- and far-field anchor-window lengths were varied from $0.5$ to $1.5~\mu\mathrm{s}$ while keeping the backscatter window length constant at 1400 time points ($7~\mu\mathrm{s}$), and the absolute fitted gradient of the cross-compensated mismatch profile was recomputed for each opposing-face pair. Across all tested anchor lengths and signal pairs, the absolute gradient remains small, with all values below $0.017$. This indicates that the cross-directional compensation method consistently suppresses the dominant monotonic depth-dependent bias and is not strongly dependent on the specific anchor-window length used.

\begin{table}[htbp]
\centering
\caption{Robustness of the cross-compensated monotonic bias metric to anchor-window length. Values represent the absolute fitted gradient of the cross-compensated mismatch profile $\Delta(z)$ for different near- and far-field anchor-window lengths.}
\label{tab:anchor_window_length}
\begin{tabular}{c c c c c c c}
\toprule
Sample & Pair & 
\multicolumn{5}{c}{Anchor-window length ($\mu$s)} \\
\cmidrule(lr){3-7}
 &  & 0.50 & 0.75 & 1.00 & 1.25 & 1.50 \\
\midrule
Big   & 1--4 & 0.0078 & 0.0085 & 0.0077 & 0.0082 & 0.0023 \\
Big   & 2--5 & 0.0070 & 0.0021 & 0.0091 & 0.0045 & 0.0020 \\
Big   & 3--6 & 0.0128 & 0.0084 & 0.0053 & 0.0038 & 0.0017 \\
Small & 1--4 & 0.0070 & 0.0064 & 0.0032 & 0.0037 & 0.0002 \\
Small & 2--5 & 0.0162 & 0.0167 & 0.0111 & 0.0111 & 0.0001 \\
Small & 3--6 & 0.0076 & 0.0051 & 0.0036 & 0.0016 & 0.0012 \\
\bottomrule
\end{tabular}
\end{table}

\section{Discussion}
\subsection{Signal Normalisation Prior to Microstructure Inference}

The proposed cross-directional compensation method should be viewed primarily as a signal normalisation step rather than a direct microstructure inversion procedure. In heterogeneous materials, depth-resolved ultrasonic backscatter measurements are affected by cumulative propagation effects that introduce systematic distance-dependent trends in the measured signals. These trends can obscure the interpretation of spatial variations in scattering behaviour associated with the underlying microstructure.

By estimating and removing the shared propagation-related baseline between opposing inspection directions, the proposed approach isolates the residual variations in the backscatter signals that arise from directional scattering behaviour. The compensated profiles therefore provide a more consistent representation of depth-resolved scattering intensity.  Hence, the proposed method serves as a preprocessing step that improves the interpretability of ultrasonic backscatter measurements, making the signals more reliable as inputs for subsequent microstructure characterisation or inversion frameworks.

\subsection{Volumetric Reconstruction of Backscatter Hotspots}

\begin{figure}
\centering
\includegraphics[width=1\textwidth]{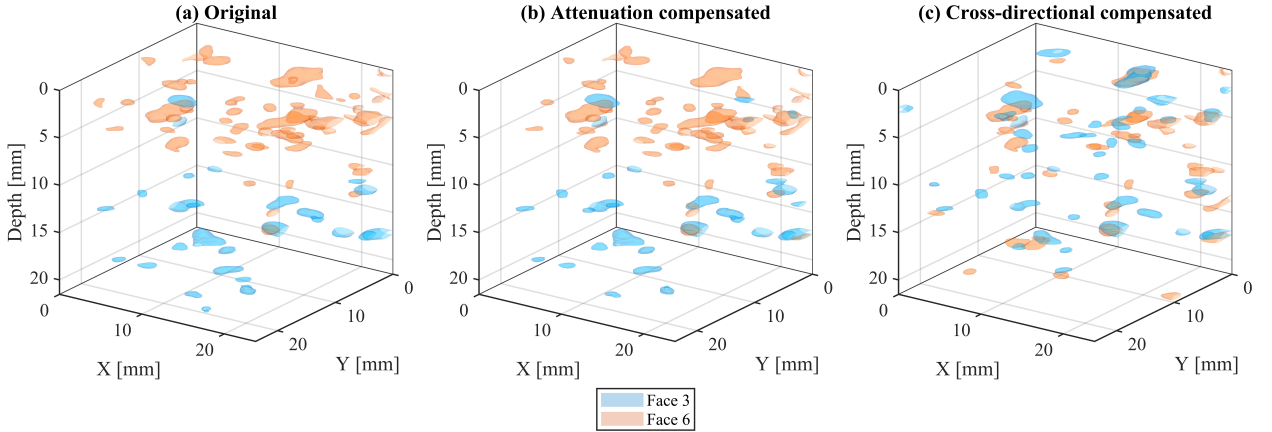}
\caption{Volumetric reconstruction of backscatter hotspots obtained from big-MTR face 3--6 dataset: (a) original, (b) attenuation-compensated, and (c) cross-compensated. Blue and orange volumes correspond to hotspots detected from face~3 and face~6 inspections, respectively. Cross-directional compensation improves the spatial alignment of hotspots across the inspection volume, indicating reduced directional propagation bias.}
\label{fig:Hotspot}
\end{figure}

To further illustrate the impact of the cross-compensation method on the spatial consistency of the backscatter response, a volumetric hotspot reconstruction was performed using the big-MTR face 3--6 backscatter dataset, as shown in Figure~\ref{fig:Hotspot}. Dominant backscatter regions were identified using a percentile-based threshold applied consistently across the (a) original, (b) attenuation-compensated, and (c) cross-compensated volumes, followed by the removal of isolated clusters smaller than 500 voxels to increase the emphasis on the dominant scattering regions for comparison.

In the original data, the hotspot distributions from the two inspection faces show substantial separation along the depth direction, reflecting cumulative energy losses due to beam spreading and scattering. Conventional attenuation compensation reduces this discrepancy but only to a limited extent. Cross-directional compensation further improves the spatial alignment, with clusters from both directions exhibiting a higher degree of overlap throughout the inspection volume.

To quantify the degree of spatial agreement, the overlap between the two hotspot volumes reconstructed from the opposing inspection directions was evaluated using the Dice coefficient \cite{Dice1945} and the Jaccard index \cite{Taha2015}. For two voxel sets $A$ and $B$, these metrics are defined as
\begin{equation}
\mathrm{Dice} = \frac{2|A \cap B|}{|A| + |B|}
\end{equation}

\begin{equation}
\mathrm{Jaccard} = \frac{|A \cap B|}{|A \cup B|}
\end{equation}
where $|A|$ and $|B|$ denote the number of voxels in each hotspot volume and $|A \cap B|$ represents the number of overlapping voxels between the two reconstructions.

The results are summarised in Table~\ref{fig:HotspotTab}, showing a substantial increase in volumetric similarity after compensation. Attenuation compensation improves the Dice coefficient from $0.039$ to $0.073$, while the proposed cross-directional compensation further increases it to $0.194$. This corresponds to an approximately $2.7$-fold improvement over attenuation compensation and an approximately $5.0$-fold improvement over the original signals. A similar trend is observed for the Jaccard index, which increases from $0.020$ for the original signals to $0.039$ after attenuation compensation, and further to $0.108$ after cross-directional compensation.

\begin{table}[htbp]
\centering
\caption{Volumetric hotspot overlap between opposing inspection directions quantified using Dice and Jaccard similarity metrics.}
\begin{tabular}{lcc}
\toprule
Method & Dice coefficient & Jaccard index \\
\midrule
Original signals & 0.039 & 0.020 \\
Attenuation compensated & 0.073 & 0.039 \\
Cross compensated & 0.194 & 0.108 \\
\bottomrule 
\end{tabular}
\label{fig:HotspotTab}
\end{table}

These results indicate that the proposed cross-compensation method provides the highest spatial consistency between the two inspection directions, supporting the removal of depth-dependent propagation bias while preserving the underlying scattering features of the material. Although the hotspot regions cannot be registered directly to the EBSD maps, their millimetre-scale extent is comparable to the macrozone length scales observed in the EBSD characterisation. The hotspot analysis is therefore interpreted as a spatial consistency assessment of the ultrasonic response, rather than as direct validation of macrozone localisation. The improved overlap after cross-compensation suggests that dominant high-backscatter regions are more consistently localised between opposing inspections, reducing the likelihood that the observed mismatch is primarily caused by direction-dependent propagation bias. This interpretation is consistent with previous experimental studies reporting stronger backscattering signals from larger macrozones~\cite{Baelde2018,Blackshire2019}.

\subsection{Assumptions and Limitations}
The proposed cross-compensation method assumes that the dominant propagation-related bias can be approximated by a smooth shared baseline between opposing inspection directions. Because the method is applied to RMS backscatter profiles averaged over the scanned region, it also assumes that the region has a sufficiently consistent through-depth propagation trend for the RMS profile to represent an area-averaged bias. The method is therefore most appropriate when opposing access is available and when the propagation bias varies slowly relative to local scattering fluctuations.

The method is not intended to recover a unique physical attenuation coefficient, nor does it remove all direction-dependent scattering effects. In cases where strong local phase scrambling, mode conversion, highly localised macrozone clusters, or strongly spatially varying propagation paths dominate the response, residual mismatch may remain after compensation. These residuals are not treated as errors, but as direction-dependent backscatter features retained by the normalisation procedure after removal of the smooth propagation-related bias.

A further limitation is that the compensated backscatter response is not calibrated directly against macrozone size, morphology, orientation, or spatial extent. The EBSD maps provide important microstructural context, but they are not registered volumetric measurements of the exact inspected region. Therefore, the residual fluctuations and volumetric hotspots are not interpreted as direct measurements or reconstructions of macrozone morphology. Instead, they are used in this study as signal-level indicators of spatial and directional backscatter consistency. A rigorous calibration of these features against the underlying macrozone field would require registered volumetric microstructure measurements or controlled numerical simulations in which the scattering distribution is known.

\section{Conclusion}
This study has presented a cross-directional compensation method for ultrasonic backscatter measurements in macrozone-containing materials, motivated by the limitations of conventional attenuation-based correction approaches under strongly anisotropic and spatially heterogeneous conditions. Experimental measurements on two macrozone-containing Ti--64 samples revealed pronounced directional differences in backscatter amplitude and decay behaviour, reflecting the presence of non-uniform macrozone distributions and significant crystallographic texture. 

By exploiting opposing inspection measurements of the same component, the proposed method estimates a shared depth-dependent propagation trend consistent with the reciprocal constraint on through-thickness transmission. Rather than enforcing pointwise symmetry, this reciprocity-based baseline reduces depth-related bias while preserving directional differences arising from heterogeneous scattering behaviour. Across the six opposing-face pairs, the mean standard deviation of the mismatch profile decreased from $0.37$ to $0.12$, while the mean absolute gradient of $\Delta(z)$ decreased from $0.171$ to $0.0067$, corresponding to reductions of approximately $67\%$ and $96\%$ relative to the original profiles. Compared to conventional attenuation compensation, the corresponding additional reductions are approximately $39\%$ and $90\%$, respectively. This demonstrates that the proposed compensation suppresses propagation-induced bias while retaining physically meaningful directional scattering variability.

The compensated backscatter RMS profiles represent the intended output of the method. These corrected signals are not interpreted here as direct indicators of macrozone size, morphology, or spatial extent. Instead, the compensation acts as a signal normalisation step that produces a more consistent and interpretable representation of depth-resolved backscatter amplitude, providing an improved input for subsequent characterisation or inversion frameworks. As a spatial consistency assessment, volumetric reconstructions of backscatter hotspots obtained from opposing inspection directions exhibited substantially improved spatial overlap after cross-directional compensation, consistent with the increased Dice and Jaccard similarity metrics observed in the volumetric analysis.

Overall, the proposed cross-directional compensation strategy offers a practical and low-assumption signal-level correction that requires no additional experimental measurements or detailed microstructural models. The method is straightforward to implement and may be extended to other highly scattering anisotropic materials where depth-dependent energy loss and directional propagation effects complicate ultrasonic backscatter-based analysis.

\section*{Funding}
This research was supported by Agency for Science, Technology and Research (A*STAR) under the National Science Scholarship (NSS-PhD). The funding body had no role in the design of the study, data collection, analysis, interpretation, or writing of the manuscript.

\bibliographystyle{unsrtnat}

\bibliography{references}

\end{document}